\documentclass[a4paper, 11pt, twoside]{article} 
\usepackage[utf8]{inputenc}
\usepackage[a4paper, top=25mm, bottom=25mm, left=25mm, right=25mm]{geometry} 
\usepackage{lmodern}
\usepackage{verbatim}

\usepackage{parskip}
\usepackage[sort&compress]{natbib}
\let\OLDthebibliography\thebibliography
\renewcommand\thebibliography[1]{
  \OLDthebibliography{#1}
  \setlength{\parskip}{3pt plus 1pt minus 1pt}
  \setlength{\itemsep}{0pt plus 0.5ex}
}

\usepackage{graphicx}
\usepackage{grffile}
\usepackage[%
    font={small,sf},
    labelfont=bf,
    format=hang,    
    format=plain,
    margin=0pt,
    width=0.8\textwidth,
]{caption}
\usepackage[list=true]{subcaption}
\graphicspath{ {../figures/} }
\usepackage{tikz}
\usetikzlibrary{calc}

\usepackage[pagebackref, breaklinks]{hyperref}

\renewcommand*{\backref}[1]{}

\usepackage{amsmath}
\usepackage{amsfonts}
\usepackage{mathtools}

\makeatletter
\newcommand*\bigcdot{\mathpalette\bigcdot@{.5}}
\newcommand*\bigcdot@[2]{\mathbin{\vcenter{\hbox{\scalebox{#2}{$\m@th#1\bullet$}}}}}
\makeatother

\usepackage{url}
\usepackage{empheq}

\usepackage{authblk}

\title{Coherent diversification in corporate technological portfolios}
\author[2]{Emanuele Pugliese}
\author[2]{Lorenzo Napolitano}
\author[2]{Andrea Zaccaria*}
\author[1]{Luciano Pietronero}
\affil[1]{\small{Department of Physics, Sapienza University of Rome, Rome, Italy}}
\affil[2]{\small{Institute for Complex Systems, Consiglio Nazionale delle Ricerche - Rome, Italy.}}
\affil[*]{\small{and.zaccaria@gmail.com}}

\begin{document}

\maketitle

\begin{abstract}
We study the relationship between firms' performance and their technological portfolios using tools borrowed from the complexity science. In particular, we ask whether the accumulation of knowledge and capabilities related to a coherent set of technologies leads firms to experience advantages in terms of productive efficiency. To this end, we analyzed both the balance sheets and the patenting activity of about 70 thousand firms that have filed at least one patent over the period 2004-2013. From this database it is possible to define a measure of the firms' coherent diversification, based on the network of technological fields, and relate it to the firms' performance in terms of labor productivity. 
Such a measure favors companies with a diversification structure comprising blocks of closely related fields over firms with the same breadth of scope, but a more scattered diversification structure. We find that the coherent diversification of firms is quantitatively related to their economic performance and captures relevant information about their productive structure. In particular, we prove on a statistical basis that a naive definition of technological diversification can explain labor productivity only as a proxy of size and coherent diversification. This approach can be used to investigate possible synergies within firms and to recommend viable partners for merging and acquisitions.
\end{abstract}

\pagebreak

\section{Introduction}

The aim of this paper is to shed light on the interactions and positive synergies that take place between corporate R\&D activities in different fields as reflected by the composition of the patent portfolios of a large sample of firms.
In particular, we show that benefits for patenting companies accrue not so much from the number of technologies in which they perform R\&D, but rather from the average size of the coherent blocks of knowledge stock in which their research activities concentrate.
As we will see, such benefits can be measured in terms of productive efficiency.
Such counter-intuitive finding is however coherent with a representation of production in which coherent knowledge blocks map to distinct internally consistent product lines.
In order to illustrate the relevance of knowledge blocks, we define a measure of coherent diversification that weighs the fields of technology contained in corporate technological portfolios based on their coherence with respect to the firm's global knowledge base.
As a result, given the same breadth of scope, this measure distinguishes companies with a diversification structure comprising blocks of closely related fields from companies with a more scattered technological portfolios. The idea is that a coherent diversification structure, being a direct reflection of production lines, leads to better economic performances. \par

While even a simpler theory of the firm would be able to capture the relevance of the technology scope for the productivity of the firm \citep{penrose1959the}, a more refined capabilities-based approach is necessary to argue that potential returns to scope are related to the nature and complementarity of the pursued technologies , and not simply their number. 
Capabilities have been originally defined as intangible assets of the firm relating to the necessary know-how enabling the effective development of production and other internal organizational processes \citep{dosi2008organizational}. 
A capability-based model of the firm can be seen as a network connecting specific technological or organizational capabilities to one or more products, thus creating heterogeneous and non trivial interactions between specific technological fields. 
Starting with \citep{teece1994understanding}, many studies have tried to exploit data on firms and products to understand the possible synergies between different \emph{products}.
A different perspective on the same concept has been championed in recent years by the literature on economic complexity, which has modeled capabilities as an invisible layer linking economic agents to the outcome of their activities \citep{hausmann2011network} and has also successfully attempted to extend the notion outside the corporate domain by applying it to nations and geographical regions in general \citep{tacchella2012new, zaccaria2014related}. 
In a way, this work lies in-between the traditional and the complexity view by modeling capabilities as a hidden layer while interpreting them as mediators between firms and their productive efforts. 
Differently from both the above approaches, however, the analysis focuses on the production of technological innovation instead of production in the traditional sense, thus applying a notion that is also close in spirit to the technological competencies proposed by \cite{granstrand1997multi}. \par

The following sections will lead up to the main exercise and the discussion of the results by first briefly describing the data we employ for the study; reviewing the relevant literature on the topic of corporate diversification while presenting an overview of prominent diversification measures; finally, we will turn to the linked concepts of relatedness and corporate coherence, which are useful stepping stones to make the link between our proposal and the existing literature explicit while highlighting its original contribution to the field.\par

\section{Data}

\subsection{Firm data} \label{sec:firmdata}

Our aim is to investigate the relation between the structure of the technological portfolios of firms and and their performance and efficiency.
To this end, we rely on AMADEUS, a commercial database maintained by Bureau van Dijk Electronic Publishing (BvD) which specializes in providing financial, administrative, and balance sheet information about (almost exclusively private) companies around the world.
In particular, AMADEUS focuses on European enterprises.
The database is comprehensive, though not exhaustive, accounting for over 20 million companies sourced from several providers using a multitude of data typically collected by public institutions \citep{ribeiro2010oecd}.
The AMADEUS sample has a straightforward connection with the patent database PATSTAT, described in the following section, because both index patent applications with the same patent identifier.
Joining the two datasets yields detailed information about almost 70 thousand firms that have filed at least one patent over the period 2004-2013 covered by our AMADEUS edition and display the balance sheet information we require concerning firm size and productivity.
A drawback associated to AMADEUS is that it concerns a set of firms which is geographically constrained to Europe. As consequence, our results draw on evidence based on companies with at least one subsidiary in Europe, which implies that only relatively large non-European firms are included.
Section \ref{sec:patdata} explains the strategy we adopt to balance the sample and limit geographical biases.

\subsection{Patents and technology codes} \label{sec:patdata}

Following an established tradition in the economic literature on innovation \citep{griliches1990patent, hall2001nber, strumsky2012using}, we proxy innovative activity with patents, a rich and growing source of information, which over the past years has benefited from cumulative data collection efforts of scholars as well as public agencies.
For the present analysis we concentrate on information concerning the set of technological fields to which inventions pertain; each field is represented by a standard code defined within the International Patent Classification (IPC), an internationally recognized hierarchical classification system maintained and constantly updated by the World International Patent Organization (WIPO) \citep{joo2010measuring}.
Apart from the obvious practical advantages of relying on standardized definitions, decomposing patents into their constituent technologies allows to consider inventions as the product of a successful recombination of variously related preexisting technologies and knowledge.
The heart of patent applications are the claims, \emph{i.e.} the part of the patent document that describes the novel aspects of the invention with respect to the relevant prior art and justifies the request for protection and, implicitly, delimiting the scope of the pursued monopoly rights.
Claims undergo individual examination by patent office officials and, if approved, are assigned one or more IPC codes relating to the technologies touched upon by the corresponding claim.

The main source of patent (and technology ) data for this paper is the EPO Worldwide Patent Statistical Database (PATSTAT), maintained by the European Patent Office (EPO), which aggregates data from national and regional patent offices and presents the information in a clean and organized fashion. 
For example, multiple patent application documents can often be referred to the same invention.
In these cases, PATSTAT collects groups of related documents into so-called patent families representing sets of patents filed in  more than one country to protect a single invention by recognizing the link between the first application --- the priority --- and later ones filed at to other patent offices \citep{martinez2010insight}. 

As we mentioned in section \ref{sec:firmdata}, the geographical coverage of AMADEUS is limited to companies with at least one European subsidiary and this could potentially over-represent Europe in general over other areas.
In order to mitigate the issue, we take into account only the subset of PATSTAT families that also responded to the more restrictive criteria defining triadic patent families \citep{dernis2004triadic}, \emph{i.e.} including an application filed at the EPO, one filed at the Japanese Patent Office (JPO), and one granted by the United States Patent Office (USPTO).
The above criteria select globally protected inventions and thus assure that only relatively large enterprises with global operations are included in the sample.
We thus gain, in exchange for some reduction in the number of observations, a more balanced sample that excludes relatively small European firms patenting only in national offices, which would otherwise be overrepresented.

The starting point of the analysis consists in decomposing the patent families with applications in a given year into the set of associated IPC codes and attributing the codes to patenting firms.
We then assign each active family a weight equal to one and split it equally among all of the unique technology-company pairs.
Every pair maps to an element of a binary matrix $M$, the value of which is computed based on the sum of the shares of active patent families attributed to the corresponding combination of technology and applicant firm (this procedure is fully described in the supplementary material).
To summarize, $M$ defines the technological portfolio embedded in the patents filed by all active firms in a target year and thus allows us to look into the structure of such portfolios, which we then relate to corporate efficiency.\par

\section{Literature review} \label{sec:literature}
This section is devoted to presenting some prominent diversification and relatedness measures proposed in the literature that are relevant to the present inquiry.

Though the empirical exercises contained in the present paper are concerned with technologies and not with products, for historical reasons the discussion of previously existing measures will always move from contributions addressing corporate productive scope and then extend to the literature concerning technological scope.
In fact, technology has come to prominence more recently in the literature than production, so that the roots and initial inspiration of most studies can be traced back to the latter stream.
This is not to say that we consider productive scope relevant only for historical reasons; on the contrary, we argue that, though the technological and productive dimensions are very different, they are strongly interconnected and they complement each other to drive the dynamics of firms.
The aim of the present section is to put into perspective the salient features of the methodology we propose while highlighting its relation to the existing contributions. 
We do not attempt an exhaustive review of the literature\footnote{See \emph{e.g.} \cite{knecht2013diversification} for a comprehensive review of the diversification measures adopted in the economics and management literature, their implications, and their theoretical foundations and \cite{} for a review of the measure of coherence.}, as it would be beyond the scope of the paper; rather, our aim is to provide a concise overview of some of the indices of diversification that have been proposed over time and to use it as the starting point to trace the evolution in the literature that has later led scholars to concentrate also on the close, albeit distinct, topics of relatedness and coherence, which are the building blocks of the approach we propose in this paper. 

\subsection{Productive and technological diversification} \label{sec:diversification}

Firm diversification and its implications have interested scholars at least since the work of \citeauthor{penrose1959the}, who has noted that the ``firm is not confined to 'given' products, but the kind of activity it moves into is usually related in some way to its existing resources'', because there are ``pools of unused productive services [\dots which,] together with the changing knowledge of management, create a productive opportunity which is unique for each firm.'' \citep{penrose1960growth}.
That essay presented strong evidence from case studies corroborating the intuition that the role played by diversification in shaping corporate opportunities is substantial.
The interest for the topic has been subsequently kept alive by several scholars (\emph{e.g.} \cite{gort1962diversif, rumelt1974strategy}), who have expanded upon the issue posed by \citeauthor{penrose1959the} and reframed the general problem in quantitative terms, extending the analysis to larger samples of firms from different industries.
Early quantitative studies concerning diversification concentrated mainly on the productive scope of companies, which was measured by accounting for the material inputs or outputs of manufacturing firms and by grouping activities together based on official industrial classifications -- \emph{e.g.} \emph{Standard Industrial Classification} (SIC) codes \citep{gort1962diversif, berry1971diversif} -- or on categories developed autonomously by the author based on prior research \citep{rumelt1974strategy, rumelt1982diversification}.
For example, \citeauthor{gort1962diversif} built his measure of diversification based on the share of company payrolls going to individual manufacturing activities, while \cite{berry1971diversif} proposed a measure that summarized the spectrum of activities of large manufacturing firms based in the USA by measuring the distribution of their output across 4-digit SIC industries and summarizing the vector of output shares with a concentration index inspired by \cite{herfindahl1950concentration}.

In addition to the wealth of theoretical contributions and industry classification attempts spurred by the widespread interest that surrounded corporate product diversification \citep[see \emph{e.g.}][for an interesting discussion]{montgomery1994corporate}, much empirical work was also devoted to understand the relation between firm performance and the number of business activities and markets entered \citep[\emph{e.g.}][]{palepu1985diversification, palich2000curvilinearity,  miller2004firms}.
However, products are not the only area in which companies show evidence of diversification and it did not escape scholarly attention that the definition of the corporate technological scope is as strategic for businesses as the decision concerning the number of markets to enter and product lines to bring to market. 
This has been thought to be especially true since the last decades of the twentieth century, which have seen the emergence of rising complexity in products and production processes \citep[\emph{e.g.}][]{rycroft1999complexity, cohen2000protecting}, increasing specialization in knowledge production \citep{pavitt1998technologies} and an accelerated pace of innovation in many industries.
This has made ``diversity particularly across technologies [\dots] no longer a choice'' \citep{fai2004technological}.

\subsection{Relatedness} \label{sec:relatedness}

\subsubsection{Product relatedness}

One of the first attempts to implicitly account for relatedness alongside diversification was made by \cite{rumelt1974strategy} and later refined in an attempt to establish a link between corporate strategy and profitability \citep{rumelt1982diversification}. 
This implied a strong change in perspective with respect to earlier papers, which concentrated on measuring industrial diversification only by the observed breadth in scope of business activities.
Such shift materialized in the growing interest towards new tools to explore corporate and industrial evolution based on measuring the \emph{distance} between the activities in which firms diversify. 
\citeauthor{rumelt1974strategy} employed a different approach with respect to his predecessors and abandoned official industry classification codes as a means to define the set of activities engaged in by firms. 
Instead, he focused on a categorical classification (not an index) of diversification strategies he elaborated himself based on the historical observation of a sample of large US industrial firms.
In particular, the author started by assessing the share of revenues due to single product lines, he then established the degree of relatedness (``absence or existence of shared facilities, common selling groups, and other tangible evidence of attempts to exploit common factors of production'') between business units , and finally assigned firms to different categories according to a composite index accounting simultaneously for product diversification and the contribution to the company's revenues of the largest group of closely related products. 
According to \cite{rumelt1982diversification}, the importance of addressing relatedness stemmed from the need to test the hypothesis, formulated based on anecdotal evidence from US manufacturing, that amidst diversified firms ``the highest levels of profitability were exhibited by those having a strategy of diversifying primarily into those areas that drew on some common core skill or resource.''
\cite{teece1994understanding} built on the intuition behind \cite{rumelt1974strategy} and embraced the view that the implications of scope for the evolution of companies and industrial structure can be better understood by including in the analysis an assessment of the overall coherence of corporate activities.
This approach reflected the more general assumption that the strategic motives behind diversification should be accounted for in order to build a taxonomy of corporate types which, in turn, could be usefully incorporated in a theory of their evolution.
To this end, they relied on a much larger sample than their predecessor by resorting to census data about US manufacturing plants.
The much larger size of the data forced them to forgo reliance on ``adjudged relatedness'' and instead employ the standard 4-digit SIC classification as the basis for the classification of the activities of individual plants.
This implied that relatedness needed to be measured instead of assumed or deduced from external information.
In particular, \cite{teece1994understanding} based the measurement on the \emph{survivor principle}, \emph{i.e.} the assumption that economic competition eventually drives inefficient organizational forms out of the market, thus promoting the co-occurrence of activities that are well integrated with one another through the reliance on complementary \emph{technological capabilities}.
The authors argued that, in virtue of the survivor principle, the data should reveal efficient combinations of activities to occur with a significantly higher frequency than one would expect as a consequence of sheer randomness.
\cite{teece1994understanding} operationalized the above reasoning by first summarizing the activity portfolios of firms in the binary matrix $C \in \{0,1\}^{F \times P}$, where non-zero values correspond to the industries in which firms operate and then using $C$ to derive the matrix of co-occurrences $J \in \mathbb{N}^{P \times P}$, such that


\begin{equation} \label{eq:cooccur}
 J^{P}_{p  p\prime} = \sum_{f}C_{f p}C_{f  p\prime} \mbox{ .}
\end{equation}

Subsequently, they derived the significant combinations of activities through a statistic ($\tau$) based on a standard t-test comparing the values of the cells of $J^{P}$ to their expected value under the null hypothesis of random diversification. Note in passing that we have added the superscript $P$ to the above notation to highlight the fact that the measures refer to products. 
The statistic $\tau$, which tells ``the degree to which the observed linkage between the two industries exceeds that which would be expected were the assignments of industries to companies simply random and leads to define'', allowed the authors to define a measure of coherence between activities for individual multi-activity firms called \emph{weighted average relatedness} 

\begin{equation} \label{eq:war}
 WAR_{p} = \frac{\sum_{p\prime \neq p} \tau_{p  p\prime}e_{ p\prime}}{\sum_{ p\prime \neq p} e_{p\prime}} \mbox{ ,}
\end{equation}

which evaluates the relatedness of each activity $p$ of a production plant to all its other activities $ p\prime \neq p$ as the average significance of their co-occurrences weighted by the share $e_{p\prime}$ of employees working in activity $ p\prime$.
A connected measure of relatedness presented in the same paper is the \emph{weighted average relatedness to neighbors} (WARN),
which measures the strength of association between an activities similarly to WAR, but focuses only on the links between closest neighbors defined by computing the maximum spanning tree associated to $J^{P}$.
\citeauthor{teece1994understanding} have shown that, as firm scope increases, WAR tends to fall -- meaning that the average distance between all the activities of a multi-activity firm grows with diversification -- while WARN tends to rise -- which indicates that the link between more highly related activities grows stronger. 
The authors thus conclude that coherence is important, since the results about WAR and WARN jointly suggest that ``if firms grow more diverse, they add activities that relate to some portion of existing activities'' \citep{teece1994understanding}.\par

\subsubsection{Technological relatedness} \label{sec:tecrel}

As is the case for diversification, corporate coherence has originally found application in the product domain, but it has been shown to be extremely meaningful in studying firm performance and evolution from a technological viewpoint \citep{engelsman1994patent, piscitello2000relatedness, leten2007technological, joo2010measuring, rigby2015technological}.
Of course, the meaning of coherence from the viewpoint of technological innovation is different from productive coherence.
Nevertheless, the concepts are also complementary in understanding firm evolution, so it is not surprising that scholars interested in technological coherence have borrowed from the toolbox of their predecessors, who had previously addressed products.
In a well-known study, \cite{breschi2003knowledge} have recovered the contribution of \cite{teece1994understanding} and built on the methodology proposed therein to investigate whether firms tend to diversify their innovative efforts in a coherent fashion by patenting in technological fields that share a common knowledge base with the technological fields in which they innovated in the past.
In analogy with \cite{teece1994understanding}, \cite{breschi2003knowledge} 
have analyzed the technological diversification of firms employing a matrix of co-occurrences

\begin{equation} \label{eq:cooccur_b}
  J_{t t\prime} = \sum_{f}M_{f t} M_{f t\prime} 
\end{equation}

akin to matrix $J^{P}$ of equation \ref{eq:cooccur} and rejected the null hypothesis of random diversification through the statistic $\tau$, which compares the number of observed co-occurrences between technologies with the expectation under a hypergeometric distribution.
Notice that matrix $J$ above is computed from the binary matrix $M$, which is similar to matrix $C$ of \cite{teece1994understanding} with the difference that it associates non-zero values to the (firm, technology code) pair corresponding to position ${M}_{ft}$ whenever firm $f$ holds a patent in field $t$.


In another interesting paper, \cite{nesta2006firm} have studied corporate knowledge coherence in the US pharmaceutical industry and show that both the scope and the coherence of the knowledge base ``contribute positively and significantly to the firm's innovative performance'', as measured by the number of patents it produces weighted by the number of citations received.
The authors further built on the work of \cite{breschi2003knowledge} and adapted it to define a measure of knowledge coherence internal to the firm.
In particular, they employed the technological counterpart of WAR defined in equation \ref{eq:war}
where the relatedness between technological fields is now weighted by the share $p_t$ of patents per field $t$ owned by the firm.
This allowed them to define firm knowledge coherence as

\begin{equation} \label{eq:coh}
 COH = \sum_{t} \Bigg( \frac{p_t}{\sum_{t} p_t} WAR_{t} \Bigg) \mbox{ .}
\end{equation}

Regressing COH against the R\&D output of the firms in their data sample -- as measured by the number of owned patents weighted by the number of citations -- they found that both knowledge coherence and knowledge scope have a positive impact on the dependent variable.

\subsubsection{The economic complexity approach}

\begin{figure}[ht]
\centering
\subcaptionbox[Capabilities mediate between countries products]{Capabilities mediate between countries and their export baskets \label{subfig:sublabel1}}
  [0.45\textwidth]{\includegraphics[width=0.45\textwidth] {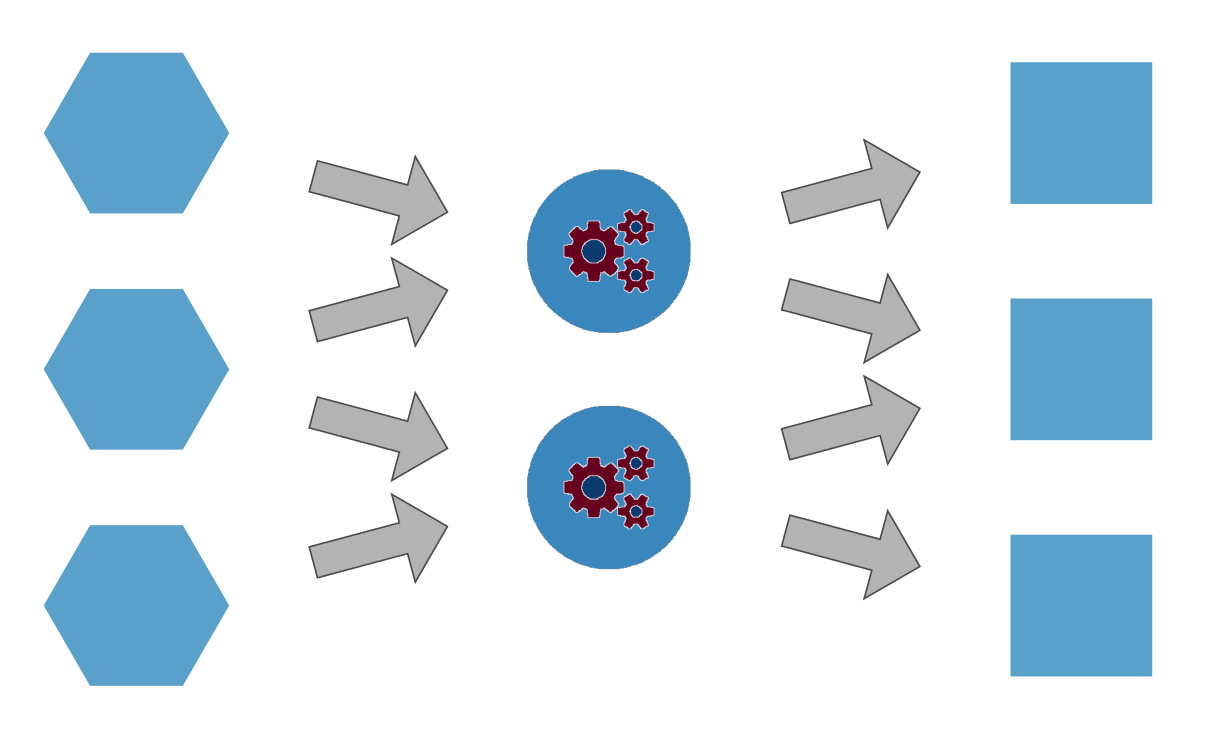}}%
\hspace{0.05\textwidth} 
\subcaptionbox[Capabilities are unobservable]{Since capabilities are unobservable, their role must be inferred from the bipartite network connecting countries to products \label{subfig:sublabel2}}
  [0.45\textwidth]{\includegraphics[width=0.45\textwidth] {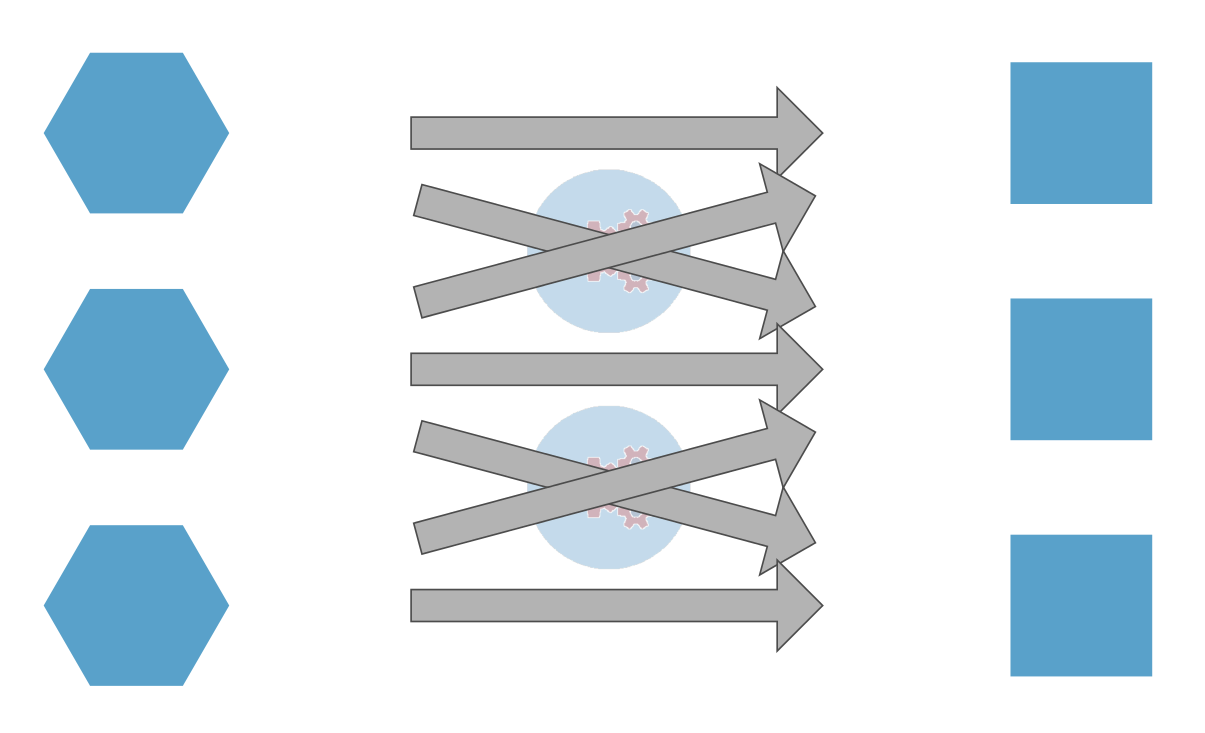}}%
\caption[\textbf{The role of intangible capabilities.}]{\textbf{The role of intangible capabilities.}}
\label{fig:capabilities}
\end{figure}

The intuition behind equation \ref{eq:cooccur} is also central in the literature on economic complexity, which in recent years has sought to explain the composition and evolution of the export baskets of nations engaging in international trade (e.g. the product space of \cite{hidalgo2007product} and the taxonomy network of \cite{zaccaria2014related}) as well as elaborate reliable predictions of the future growth trajectories of national economies \citep{hidalgo2007product, tacchella2012new}. 
The assumption underlying all the above contributions is that the patterns of competitive advantage revealed by national export baskets are the result of intangible country-level capabilities (see figure \ref{fig:capabilities}), which must be possessed and combined effectively in order to acquire the necessary strength to thrive in global competition.
These approaches aim to reconstruct a network linking products based on the similarity of the sets of capabilities required to produce them efficiently. 
Consequently, if a nation alone has a revealed competitive advantage in exporting a given good, one can infer that the nation in question possesses the needed combination of capabilities.
To this end, one can define a binary matrix in which the generic element $M_{cp}$ takes value one if country $c$ has a revealed comparative advantage \citep{balassa1965trade} in exporting product $p$.
The measure of proximity $\phi$, which is key in defining the product space of \cite{hidalgo2007product}, reads

\begin{equation} \label{eq:hidalgo_prox}
 \phi_{p p\prime} = \min \bigg( \frac{\sum_{c}M_{cp}M_{c p\prime}}{u_p}, \frac{\sum_{c}M_{c p}M_{c p\prime}}{u_{p\prime}} \bigg) \mbox{,}
\end{equation}

where $u_p \equiv \sum_c M_{c p}$ is the ubiquity of product $p$, \emph{i.e} the number of countries which export it. The proximity represents the empirical counterpart of the conditioned probability to export a product, given the export of another product.

Similarly, the formula for the taxonomy network proposed by \cite{zaccaria2014related} can be represented in terms of $M$ by defining the matrix $B \in \mathbb{N}^{P \times P}$

\begin{equation} \label{eq:zac_tax}
 B_{p p\prime} = \frac{1}{\max(u_p,u_{p\prime})} \sum_{c} \frac{ M_{cp}M_{c p\prime} } {d_c} \mbox{,}
\end{equation}

where $d_{c} \equiv \sum_p M_{c p}$ is the diversification of country $c$, \emph{i.e} the number of products it exports. From a probabilistic point of view, here the frequency of a product's occurrence is not only conditioned to the presence of another product but also evaluated with respect to a random binomial case, which would have an frequency equal to $d_c/P$ (the constant factor P is usually neglected). 
Equation \ref{eq:zac_tax} can be also interpreted, following \cite{zhou2007bipartite}, as the probability to go from a product to the other performing a random walk defined on the tripartite product-country-product network.

It is worth noting that, within the general framework of relatedness and corporate coherence, the motives guiding the authors of different contributions have been quite heterogeneous. 
On one hand, the prevalent aim of some studies has been to define a taxonomy of corporate diversification strategies and organizational structures from which to deduce the dynamic properties of industries \citep{teece1994understanding, bottazzi2010measuring}. 
On the other hand, a related stream of articles has focused on the nexus between differentiation, coherence, and performance \citep{rumelt1982diversification, breschi2003knowledge, nesta2006firm} with the aim of inferring the implications of diversification strategies for performance at the micro level.
The present paper adds to the broad literature on corporate coherence in technological portfolios by proposing a firm-level measure of coherent diversification that is inspired to the capabilities-based approach proposed by \cite{teece1994understanding} and uses the formulation introduced by \cite{zaccaria2014related}.

\section{Coherent diversification} \label{sec:measure}

\begin{figure}[tb]
\centering
\includegraphics[width=0.95\textwidth]{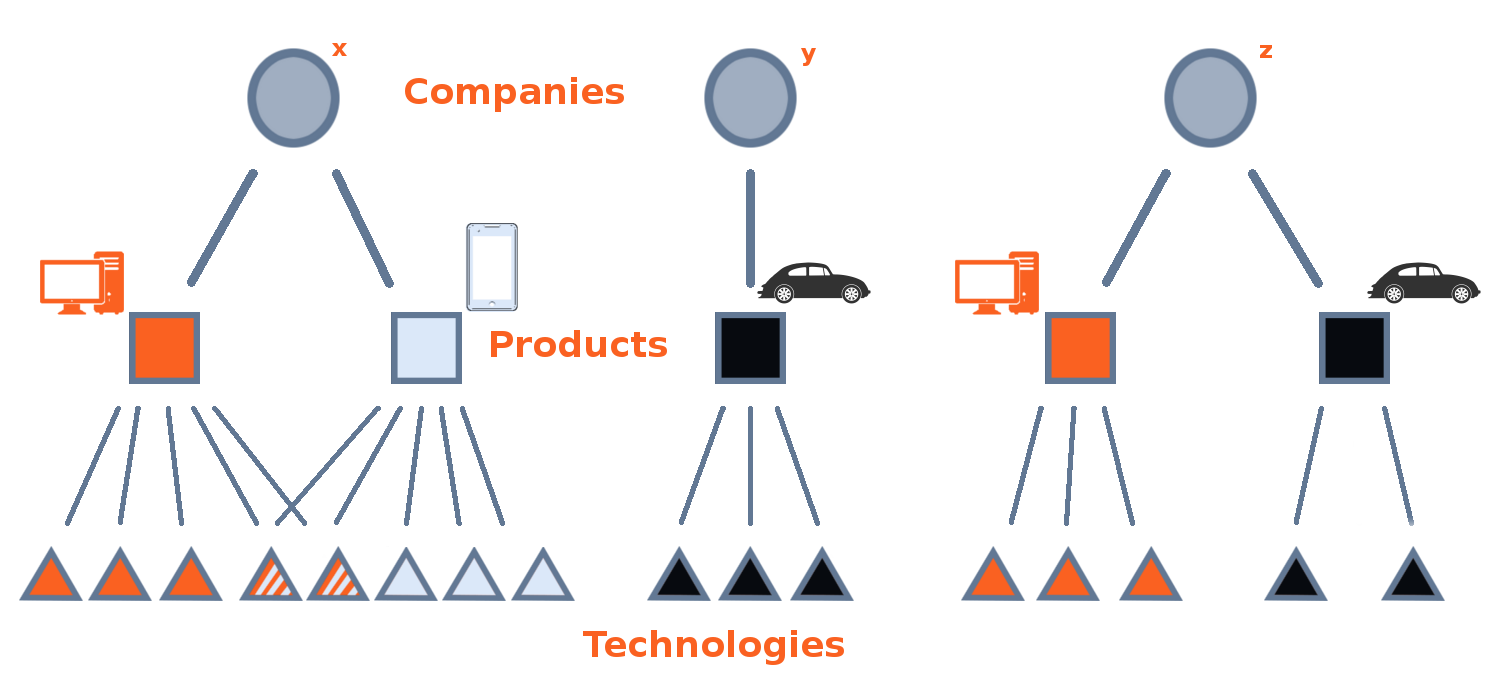}
\caption[\textbf{Corporate technological portfolios conceal information about feasible output baskets.}]{\textbf{Corporate technological portfolios conceal information about feasible output baskets.} Modeling technological portfolios to get a glimpse of the structure underlying product baskets is an operationally similar task to the one undertaken by scholars who aim to understand the relevance of intangible capabilities from the composition of the output mix produced by agents (see figure \ref{fig:latentProd}). Notice however that the two tasks are conceptually different as suggested by the fact that in figure \ref{fig:capabilities} capabilities are the actual mediators between economic agents and their output, while here products are the hidden layer.}
\label{fig:latentProd}
\end{figure}

The contribution of this paper lies at the intersection between the literature on corporate coherence \citep[especially][]{nesta2006firm} and the contributions to the economic complexity literature.
In particular, our aim is to transpose the definition of relatedness proposed by \cite{zaccaria2014related} at the firm level and apply their proposed measure (equation \ref{eq:zac_tax}) to corporate patent portfolios in order to uncover the structure of the underlying \emph{technology space}. 
This will serve as a stepping stone to elaborate a measure of the coherent technological diversification of the firms in our dataset, which we can use to examine the relation it bears with performance.
Finally, the analysis we propose also follows in the footsteps of \cite{teece1994understanding} in that we analyze the relevance of relatedness in firm diversification by building on the premise that firm-level technological capabilities are the drivers of successful diversification.
At the same time, it differs from \cite{teece1994understanding} in two ways: i) it is centered on the analysis of corporate technology portfolios arising from of capabilities related to the production of technological advancement instead of focusing on the traditional notion of capabilities, which concern the know-how involved in the creation and development of products, and ii) adopts the different formulation of co-occurencies used in \cite{zaccaria2014related}. 

\begin{figure}
  \centering
  \subcaptionbox[Matrix $M$]{Matrix $M$ \label{subfig:matrices_c}. The circles represent the firms and the triangles represent the technological fields comprising their technological portfolios.} 
  [0.45\textwidth]{\includegraphics[width=0.45\textwidth]{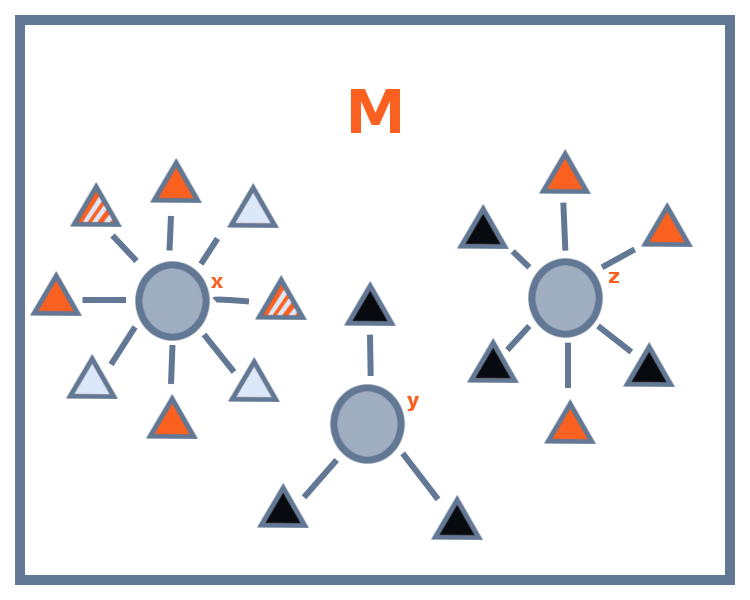}}%
  \hspace{0.05\textwidth}
  \subcaptionbox[Matrix $B$]{Matrix $B$ \label{subfig:matrices_b}. The triangular nodes in the graph correspond to a technological fields and are colored to highlight proximity to the more frequently co-occurring (thus more related) technologies hey are connected to.}
  [0.45\textwidth]{\includegraphics[width=0.45\textwidth]{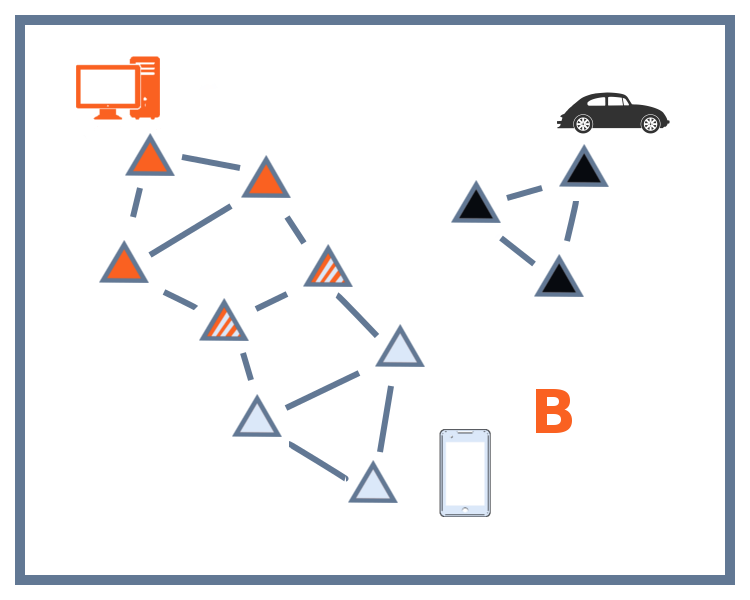}}%
\caption[\textbf{Basic Matrices.}]{\textbf{Basic Matrices.} }
\label{fig:matrices}
\end{figure}

As illustrated by figure \ref{fig:latentProd} and its comparison to figure \ref{fig:capabilities}, modeling technological portfolios to get a glimpse of the structure underlying product baskets is an operationally similar task to the one undertaken by scholars who aim to understand the relevance of intangible capabilities from the composition of the output mix produced by agents.
Conceptually, however, the two endeavors are quite different.
Notice in fact that in figure \ref{fig:capabilities} capabilities are the actual mediators between economic agents and their output.
Figure \ref{fig:latentProd} instead depicts products as the hidden layer, but it would be wrong to deduce from that products mediate between agents and technological fields, because it would amount to assume that production is instrumental to R\&D, while the relation clearly must go in the opposite direction.
Moreover, figure \ref{fig:latentProd} illustrates our view of the possible coherent structure of technological portfolios. 
The company on the left produces computers and smartphones. Some of the underlying technologies are used only for a specific products, and we color them with the same color (orange or light blue). 
However, since the two products are highly related from a technological point of view, we can assume that many of the capabilities needed to manufacture one product will be useful also for the other product. 
This situation will characterize also patenting activities, where technological fields will sometimes be shared between product lines (see the striped technologies in figure \ref{fig:latentProd}). 
In this view, the coherent company \emph{par excellence} is the one in the middle of figure \ref{fig:latentProd}, which is specialized in a single product and thus needs only the technologies related to the business activity. 
The company on the right, on the contrary, produces unrelated products; this will result in an incoherent technological portfolio. 
In what follows, we will test if the performance of a company is related not only to the diversification but also to the coherence of its technological capabilities. 
To this end we will introduce a formula to represent such features of the technological portfolios and quantify the so-called \textit{coherent diversification}.
\par

\begin{figure}[tb]
\centering
\includegraphics[width=0.95\textwidth]{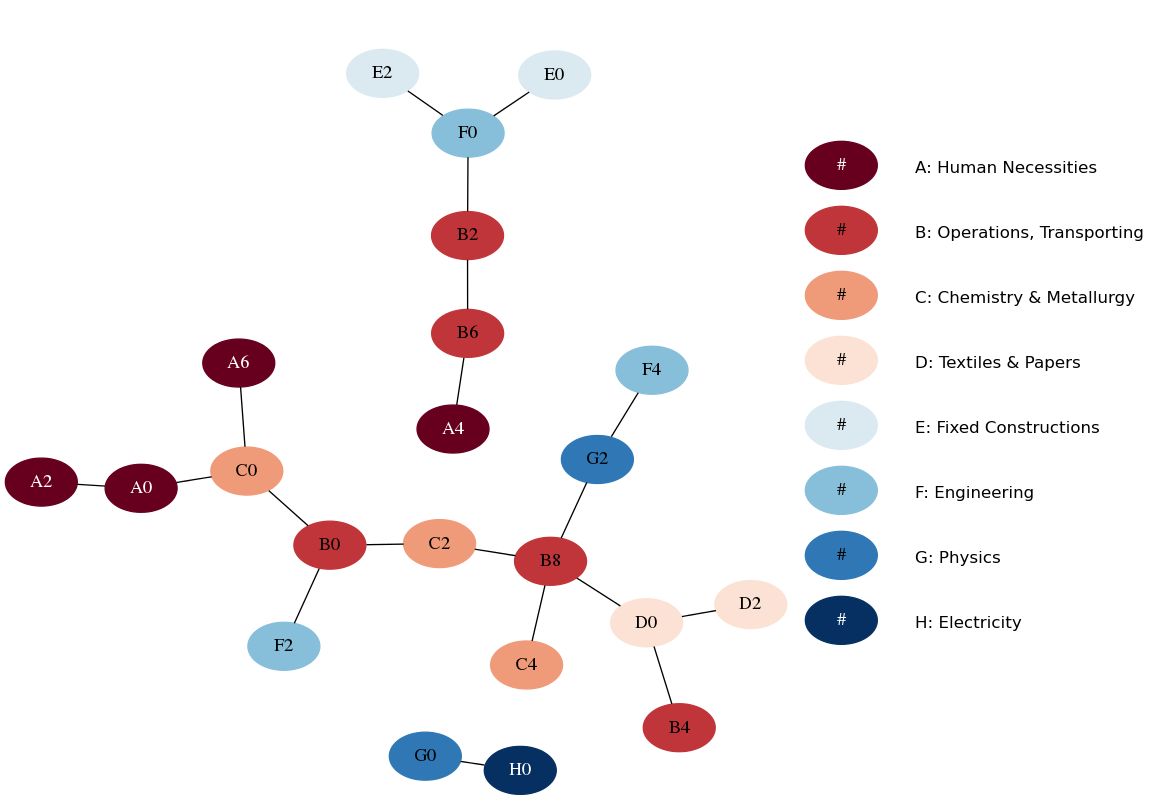}
\caption[\textbf{Minimum spanning tree of ${B}$.}]{\textbf{Minimum spanning tree of ${B}$.} The nodes in the graph represent IPC subsections and are colored according to the section they belong to.
Each node is connected to the technological field which is linked to it with the highest weight.}
\label{fig:bhat_tree}
\end{figure}

The basic data we need to define coherent diversification in corporate technological portfolios is the matrix $M$ defined in section \ref{sec:patdata} and further discussed in section \ref{sec:relatedness}. 
This matrix represents a bipartite network linking companies to the technological fields in which they are active innovators. 
For this study we perform a yearly analysis and select for each period the triadic patent families in which the firms have a stake as owners. We point out that the specific results presented below refer to the data for 2011, the most recent year for which we trust the data coverage to be reasonably complete; the results are however robust and hold also for previous time periods. 
A stylized graphical representation of the bipartite companies-technologies network, whose adjacency matrix is $M$, is depicted in figure \ref{subfig:matrices_c}.
\par

\begin{figure}[tb]
\centering
\includegraphics[width=0.95\textwidth]{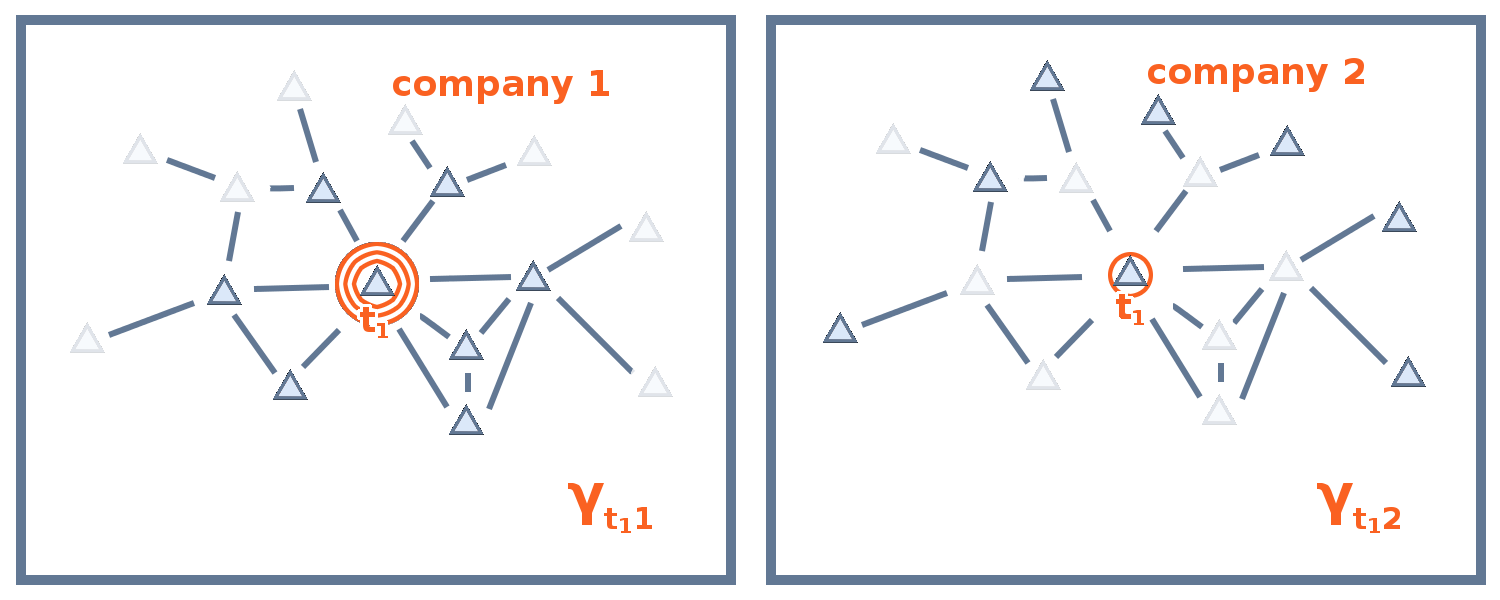}
\caption[\textbf{Illustration of $\mathbf{\gamma}$.}]{\textbf{Illustration of $\mathbf{\gamma}$} for a generic technology $t_1$ and two firms (1 and 2), depicted respectively in the left and right panels. In both panels, the graph represents the binary $B$ of figure \ref{subfig:matrices_b}: the opaque triangles stand for technological fields in which the associated firm holds patents. Both firms are diversified in the same number of technological fields. However, those of the firm 1 are connected within $B$ forming a unique block; on he contrary those of the firm 2 are scattered through the graph. As a consequence, technology $t_1$ highly coherent in firm 1 and not in firm 2.}
\label{fig:gamma}
\end{figure}

In order to define coherent diversification, we first need a measure of technological relatedness. To this end, we use the matrix $B$ of equation \ref{eq:zac_tax}, which we redefine to account for firm and technologies, obtaining

\begin{equation} \label{eq:zac_firm_rel}
 {B}_{t t\prime} = \frac{1}{\max(u_t,u_{t\prime})} \sum_{f} \frac{ {M}_{ft} {M}_{f t\prime} } {d_f} \mbox{.}
\end{equation}

The matrix ${B}$ can be interpreted as the adjacency matrix of a monopartite network of technologies like the one represented in figure \ref{subfig:matrices_b}. 
Each of the triangular nodes in the graph corresponds to a technological field and is colored to highlight its proximity to the more frequently co-occurring (thus more related) technologies to which it is linked. 
The figure shows that $B$ has embeds the notion that specific combinations of technologies concur to generate products, even though it is not possible to establish the correspondence between the technology and the production domains.
Moreover, $B$ depicts the technological space as a whole, but holds no information about the firms (the circles in figure \ref{subfig:matrices_c} representing the matrix $M$) whose technological portfolios were used to compute it. \par
Figure \ref{fig:bhat_tree} shows the minimum spanning tree defined from the empirical data used to compute ${B}$; by construction, each node represents a technological field and is connected field with which it shares the heaviest link. 
The nodes in the graph represent IPC subsections and are colored according to the section they belong to. 
The color pattern of the graph highlights a tendency of broad technological fields to connect with similar ones much less frequently than to relatively distant ones, which suggests that mixing of different fields is far from rare. 

\begin{figure}[tb]
\centering
\includegraphics[width=0.95\textwidth]{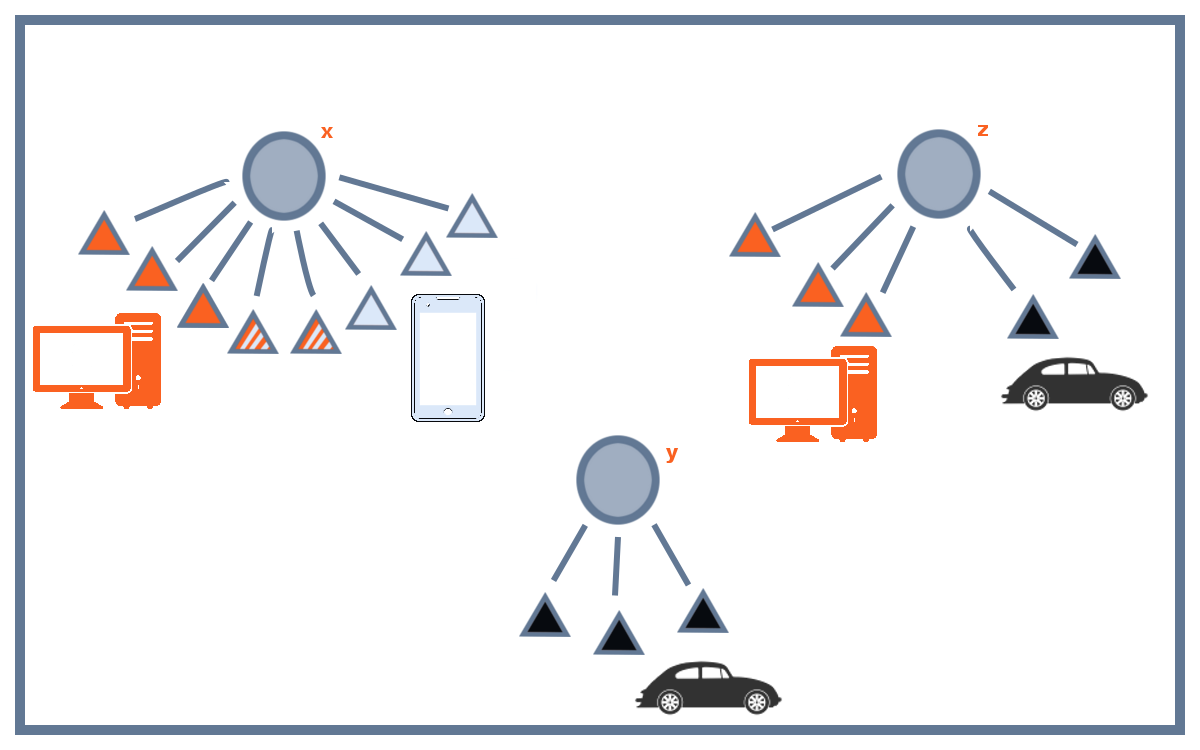}
\caption[\textbf{Illustration of $\mathbf{\Gamma}$}]{\textbf{Illustration of $\mathbf{\Gamma}$}, which can be interpreted as a reweighing of the diversification structure of firms. $\mathbf{\Gamma}$ highlights the blocks of connected technologies. In principle, it has a correspondence with the corporate product basket, however the information about the map connecting what a firm knows and what it produces remains hidden beneath the surface.}
\label{fig:Gamma}
\end{figure}

In order to combine the general structure of technology relatedness with firm-specific information, we first need to measure for each company the coherence between all of the technologies in which it holds patents. 
Figure \ref{fig:gamma} qualitatively illustrates such measure for a generic technology $t_1$ and two toy -- 1 and 2 -- depicted respectively in the left and right panels.
In both panels, the network structure connecting the triangles in the background represents a simplified (binary) illustration of $B$; the opaque triangles stand for technological fields contained in the patent portfolio of each firm, while the transparent triangles represent technological fields in which the same firm has not filed patents during the time period covered by $M$.
Notice that both firms are equally diversified because they both have patents covering the same number (eight) of technological fields.
The glaring difference between firm 1 and firm 2 resides in their diversification structures. 
In particular, the technological fields of the first company are all connected within $B$ and form a unique block, while the technologies of the second are scattered in $B$.
As a consequence, technology $t_1$, which is owned by both firms, has a high intra-firm coherence within firm 1, but attains a low score in firm 2.
In reality, the linkages we measure at each step of the analysis between companies and technology fields are not binary but, rather, weighted and we must keep this into account in the analytical definition of coherence, that will depend on both technology $i$ and the surrounding technological basket of firm $f$. 
We define the intra-firm coherence of technologies by the rectangular matrix $\gamma \in \mathbb{N}^{F \times T}$ where

\begin{equation} \label{eq:gamma}
 \gamma_{ft} =\sum_{t\prime} {B}_{t t\prime} {M}_{f t\prime} \mbox{ ,}
\end{equation}

the analytical counterpart of figure \ref{fig:gamma}.
\par

Finally, it is possible to define an index of corporate technological coherence, that we call \textit{coherent diversification}, by aggregating within each firm the information about the intra-firm coherence of all the technological fields in which it holds patents.
As schematically represented by figure \ref{fig:Gamma}, this can be interpreted as a reweighing of the diversification structure of firms, which highlights the connected technologies and in principle has a correspondence with the corporate product basket, though the information about the explicit map connecting what a firm knows and what it produces remains under the surface. \par

In formula, we define the firm-specific index of coherent diversification $\Gamma \in \mathbb{R}^{F}$ as

\begin{equation} \label{eq:Gamma}
 \Gamma_{f} =\frac{\sum_{t} {M}_{ft}\gamma_{f_t}}{d_f} \mbox{ ,}
\end{equation}

where $d_f \equiv \sum_{t}M_{ft}$ is the diversification of firm $f$. In practice, $\Gamma$ computes the average size of the coherent technology blocks comprising the  technological portfolio of each company.
\subsection{Toy examples}

\begin{figure}[tb]
\centering
\includegraphics[width=0.85\textwidth]{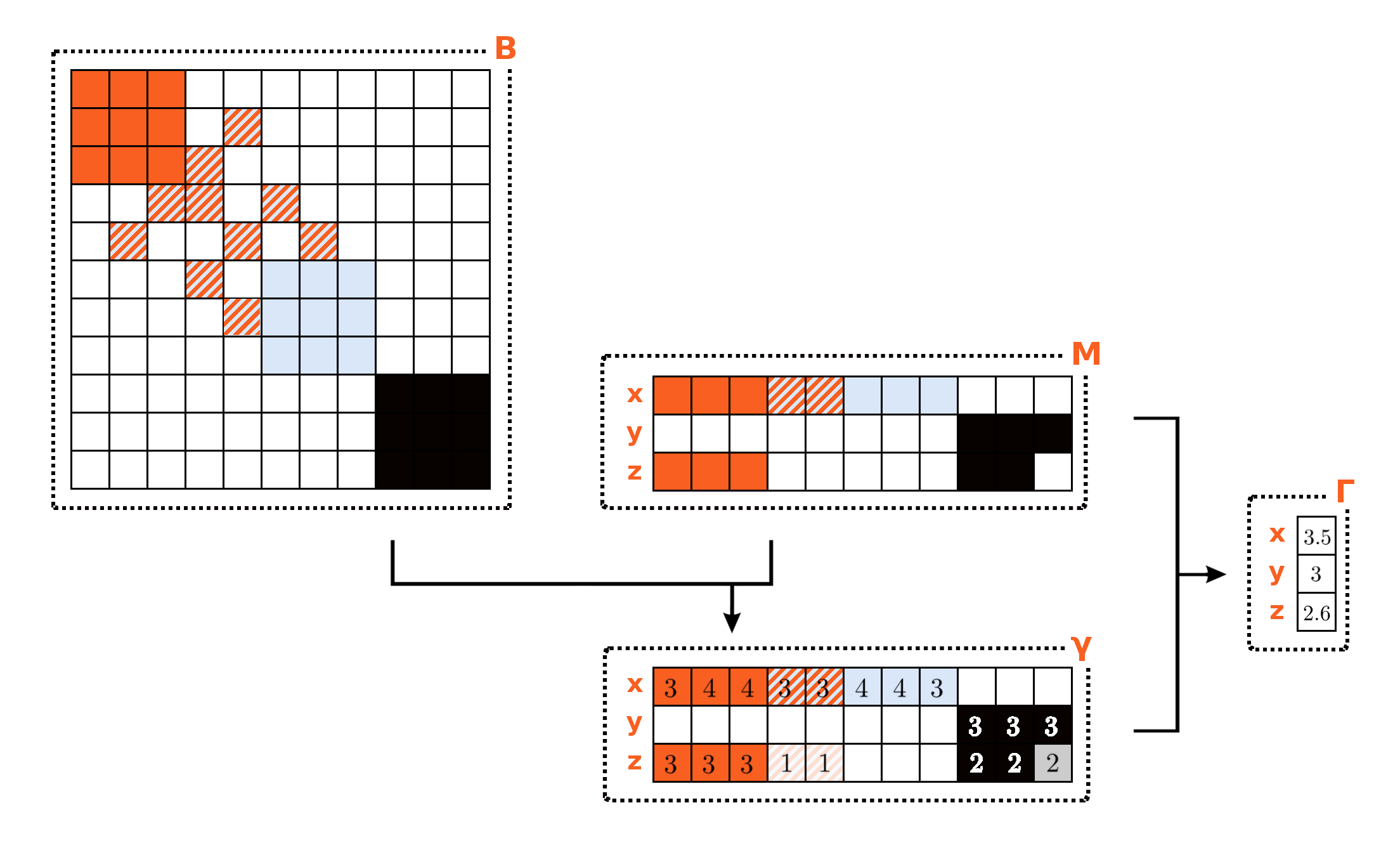}
\caption[\textbf{Toy example}]{\textbf{Toy example.} A graphical representation of the matrices ${B}$ from ${M}$ relative to the toy example discussed in the text.}
\label{fig:toy}
\end{figure}

A simple example will help explain how this framework rewards diversification only if it defines a coherent portfolio. 
Suppose that company $f$ owns two close technologies, such that $B_{t t\prime}=1 \quad \forall (t, t\prime) = \{1,2\}$. 
Straightforward calculations lead to $\Gamma_f=1+1=2$: the coherent diversification is equal to the standard definition of diversification. 
On the contrary, if these two technologies are not related, $B_{t t\prime}=\delta_{t t\prime}$, and we will have $\Gamma_f=\frac{1}{2}(1+1)=1$: in this case the lack of coherence averages out the diversification.

In order to further clarify the economic meaning of the coherent diversification and its relation with production lines, we present a simple calculation based on the illustrations shown in this section. 
Let us start from the situation depicted in figure \ref{fig:latentProd}. 
We have three companies: the first one (company x) has two production lines (computers and smartphones) and its portfolio contains eight technologies, of which three are purely related to computers, three are necessary for smartphones, and  two are useful for both products; the second company, y, is instead specialized in cars and controls three technologies related to this production line; finally, the third company, z, has two unrelated production lines, computers and cars, relying respectively on groups of three and two technologies. 
The associated ${M}$ matrix is depicted at the top right of figure \ref{fig:toy}.
In order to compute the coherence of these technological portfolios we need a measure of distance, ${B}$. 
In this example, we do not compute ${B}$ from ${M}$ like we will do for the real case; on the contrary, we suppose that the three company live in a technological space defined by other companies that are not individually included in the example. 
In particular, we take the technological network depicted in \ref{subfig:matrices_b}, whose adjacency matrix is represented in the top left of figure \ref{fig:toy}.
The technologies related to cars are homogeneous (i.e., fully connected) and independent from technologies used for their production lines (i.e., there are no off diagonal elements connecting them to other technologies), forming a single unitary block. 
On the contrary, computer and smartphone technologies are homogeneous but mildly related through the two off-diagonal technologies (the fourth and the fifth in the first row of $M$). 
Note that here we have a binary matrix, but in general the elements of ${B}$ can have any value.

Let us now compute the intra-firm coherence of technologies, that is, the enhancement technology $t$ gets thanks to the fact of being in the portfolio of company $f$. Applying equation \ref{eq:gamma} we obtain 
the bottom matrix of figure \ref{fig:toy}.
In this simple case, the matrix just counts the neighbors of a technology that are owned by the company. 
Notice that the block of car technologies is more coherent in firm y than in firm z, since they own 3 and 2 technologies in that block respectively. 
Finally, using equation \ref{eq:Gamma}, we can compute the coherent diversification of the three companies. 
For company y we obtain $\Gamma=3$. 
In this simple case, the coherent diversification is simply the average number of technologies used for each production line. 
Such interpretation is a zero order approximation, which turns to be exact only for independent and homogeneous production lines. 
Let us now consider company x. 
In this case, the enhancement due to the close technologies is stronger, as one can notice looking at the first row of the $\gamma$ matrix; averaging over the owned technologies, one obtains $\Gamma=3.5$. 
Finally, company z has $\Gamma=2.6$. 
This can be interpreted as a weighted average over the production lines: the first production line (computers) has three technologies, all with an intra-firm coherence equal to three, while the second production line (cars) can use only two technologies, and this implies a lower coherence, equal to two. 
In order to compute $\Gamma$ we weigh the coherences with the relative number of technologies used for each product: $\frac{3}{5}\times 3 + \frac{2}{5}\times 2=2.6$.   

\section{Results}

We now test the measure of firm coherence $\Gamma$ defined in the previous section \ref{sec:measure} by correlating it with an index of firm efficiency.
In fact, if our hypothesis that innovating in related technological fields is conducive to the development of an effective mix of firm-level capabilities, which is in turn reflected in production, then this should correlate with firm performance.

\begin{figure}[ht]
\centering
\includegraphics[width=0.95\textwidth]{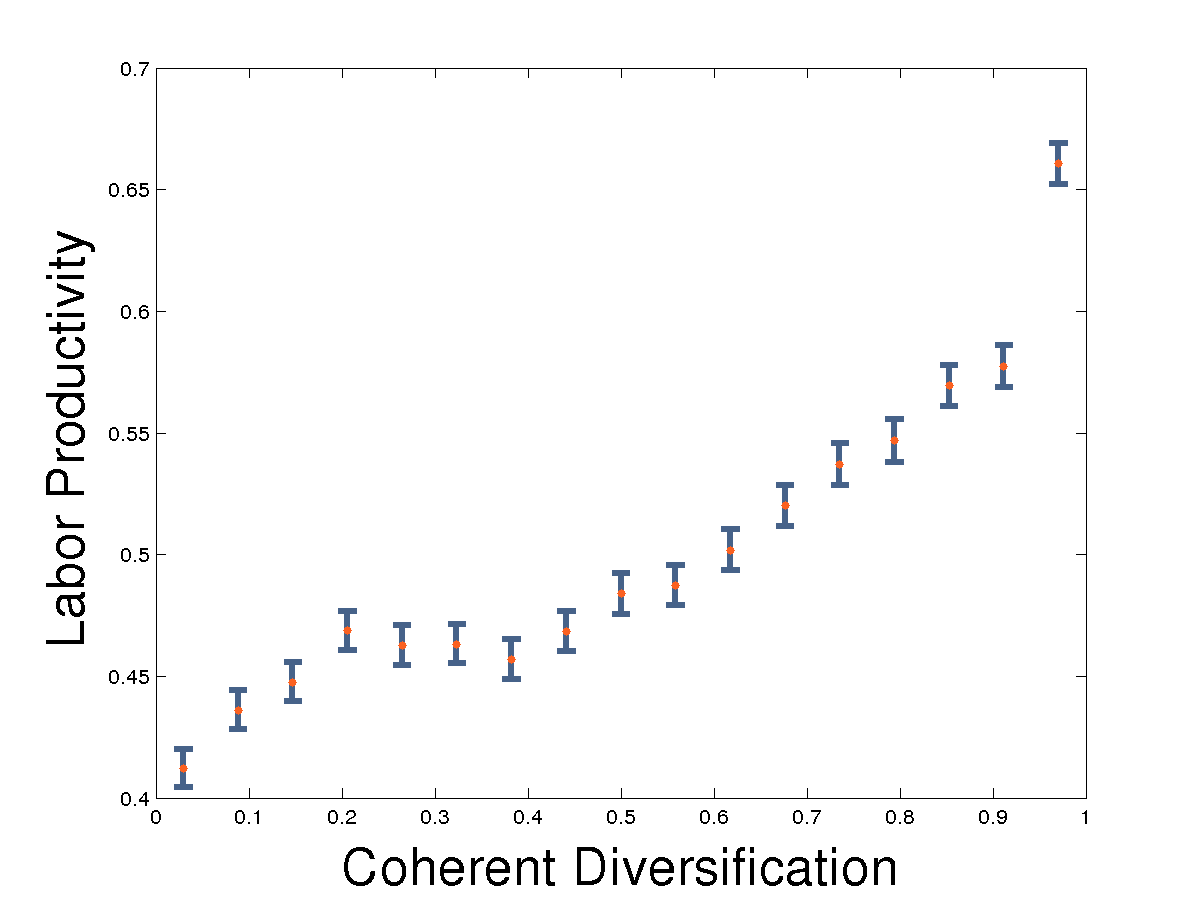}
\caption[\textbf{Coherent diversification VS labor productivity.}]{\textbf{Coherent diversification VS labor productivity.} The graph plots the binned values of related diversification ($\Gamma$) of the firms in our sample against the intra-bin quantiles of labor productivity. The clear positive association between $\Gamma$ and labor productivity suggests that coherent diversification of technological portfolios captures relevant information about the corporate productive structure.}
\label{fig:correlate}
\end{figure}

The first test is illustrated in figure \ref{fig:correlate}, which plots the binned values of $\Gamma$ against the intra-bin quantiles of labor productivity (measured as value added over employees) for the firms in our sample. 
The plot shows a clear positive association, providing preliminary evidence that our measure of coherent diversification of technological portfolios captures relevant information about the productive structure of the firms. 
As a further test of the ability of $\Gamma$ to capture a relevant aspect of corporate productive efficiency, we regress it against labor productivity.
The results of the least squares regressions, which are summarized in table \ref{tab:reg}, further confirm the intuition conveyed by figure \ref{fig:correlate}.
The coefficient associated to coherent diversification remains positive and significant in all regressions, even when we add firm size (measured by total assets) and diversification (\emph{i.e.} the number of technology codes in the firm's patent portfolio) as controls.
Moreover, though simple diversification is statistically significant if used alone, it loses explanatory power when used in the same model as $\Gamma$.
This is particularly interesting, because it suggests that the number of connected technologies within a company's technological knowledge portfolio, whose relation is quantified by our measure of coherence, is more relevant than the raw number of technological fields in which the company innovates.
In particular, the fact that the statistical significance of diversification -- as measured by the number of technologies comprising firm technological portfolios -- vanishes once coherent diversification is added to the set of regressors suggests that the former can be considered a proxy for the latter.
Our findings thus suggest that what firms know is relevant to what they produce and that the internal consistency of their knowledge stock is even more relevant than its the sheer scope. 

\setlength\arraycolsep{3pt}
\def\arraystretch{0.5}
\begin{table}[ht]
  \centering 
  \begin{tabular}{ r c c c c }
  VARIABLES   & (  0 )     & (  1 )     & (  2 )     & (  3 )      \\ 
  \hline			
  &&&\\
         Size & 0.079*** & 0.079*** &            & 0.081***  \\
              & \footnotesize{(0.023)}  & \footnotesize{(0.008)}  &            & \footnotesize{(0.008)}   \\ 
              &&&\\
  Diversification & 0.010    &            &            & 0.074***  \\
              & \footnotesize{(0.045)}  &            &            & \footnotesize{(0.009)}   \\
              &&&\\
Coherent Div. & 0.136*** & 0.154*** & 0.200*** &             \\
              & \footnotesize{(0.045)}  & \footnotesize{(0.017)}  & \footnotesize{(0.016)}  &             \\
              &&&\\
  \hline
  &&&\\
        $R^2$ & 0.063    & 0.062    & 0.040    & 0.060     \\
    
  \end{tabular}
  \caption{Regressions of labor productivity against coherent diversification, diversification, and size}
  \label{tab:reg}
\end{table}

\begin{figure}[ht]
\centering
\includegraphics[width=0.95\textwidth]{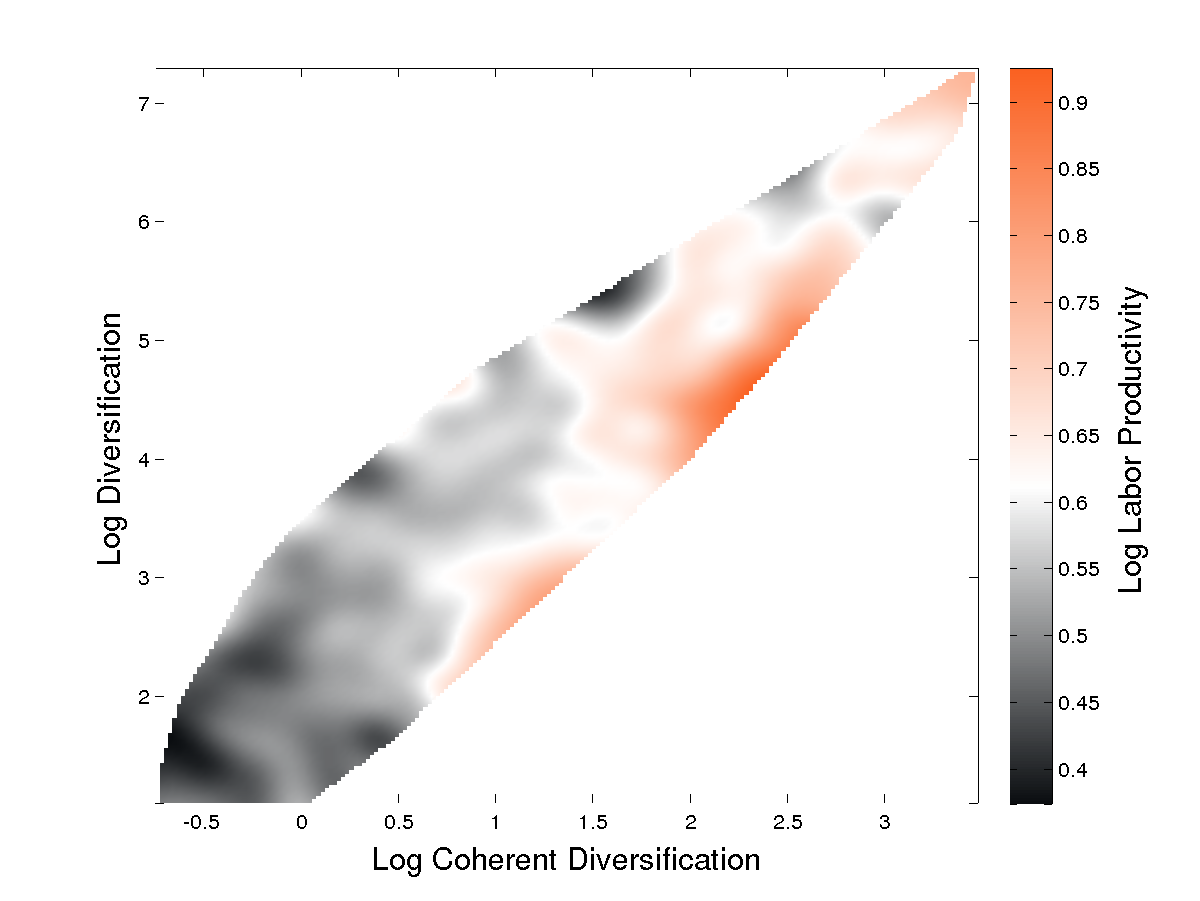}
\caption{\textbf{Labor productivity as a function of Diversification and Coherent Diversification.} A graphical representation of what we pointed out in Table \ref{tab:reg}: diversification loses its explanatory power in favor of coherent diversification when both are considered. Notice that, given a fixed value of diversification, labor productivity tends to increase with coherent diversification (i.e., from left to right, considering horizontal slices), while the opposite does not hold.}
\label{fig:divcoediv}
\end{figure}

The results shown in \ref{tab:reg} can be represented by means of a two dimensional plot, in which we consider labor productivity as a function of both diversification and coherent diversification. 
In figure \ref{fig:divcoediv} we use these two variables to aggregate the firms into square areas colored based on their ranking in labor productivity. 
As expected, there is a strong correlation between coherent and standard diversification, which leads to the presence of white (empty) squares away from the main diagonal. 
More interestingly, coherent diversification has more explanatory power with respect to the standard diversification: on average, horizontal slices exhibit a stronger gradient in labor productivity than vertical slices.

\begin{figure}[ht]
\centering
\includegraphics[width=0.95\textwidth]{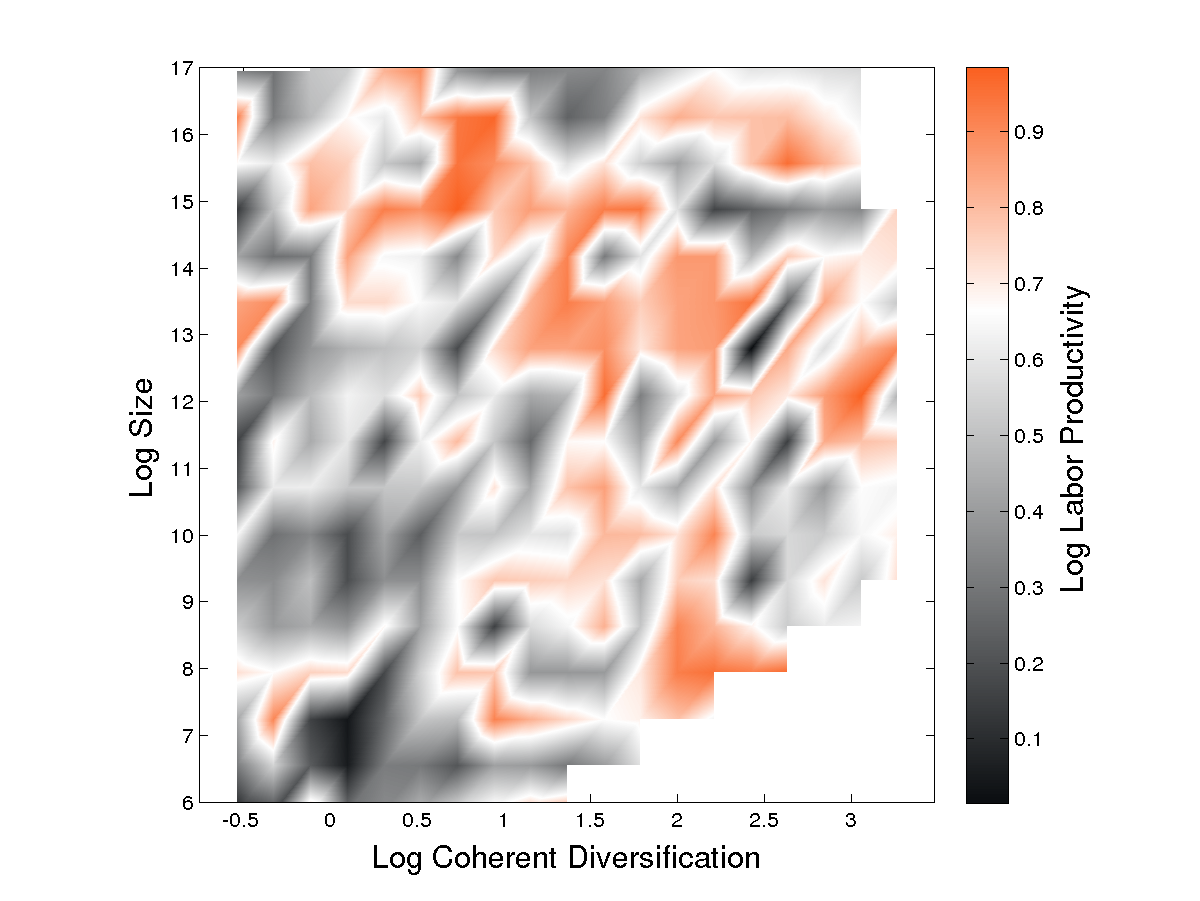}
\caption{\textbf{Labor productivity as a function of Size and Coherent Diversification.} Coherent diversification gives complementary information about firms performance.}
\label{fig:sizecoediv}
\end{figure}

We conclude this section by analyzing the role played by firms size. 
Similarly to the previous exhibit, in Fig.\ref{fig:sizecoediv} we plot labor productivity as a function of size and coherent diversification. 
The two variables are clearly complementary: on average, large size or large coherent diversification are associated with larger labor productivity, and the same holds for linear combinations of the two. 
Obviously, this is true on average, and a large degree of heterogeneity is present. 
However, the comparison of figure \ref{fig:sizecoediv} with figure \ref{fig:divcoediv} allows us to conclude that the effect on labor productivity of standard diversification depends on its correlation with size and coherent diversification, which thus represents a better framework to discuss the effects of the structure of technological corporate portfolios on firms performance. 

\section{Relevance of scale}

In this section we briefly discuss the relevance of data resolution for the results of the empirical analysis. 
Even though in this paper we have dealt with corporate patent portfolios, and have thus focused our attention into individual economic agents instead of geographical regions, there is still room for the scaling of the data to have an effect on the results.
The most straightforward way through which one could explore this effect would be to vary the coarseness of the technological classification employed to define $M$.
Though this exercise might lead to interesting explorations, we would like to concentrate on a more subtle channel through which the effect can be transmitted, namely the geographical scale used to define $B$.
Notice in fact that when $B$ enters equation \ref{eq:gamma} its size must match the number of columns of $M$, \emph{i.e.} it must represent relatedness at the same technological scale at which the technological portfolios of the firms are defined.

\begin{figure}[ht]
\centering
\includegraphics[width=0.9\textwidth]{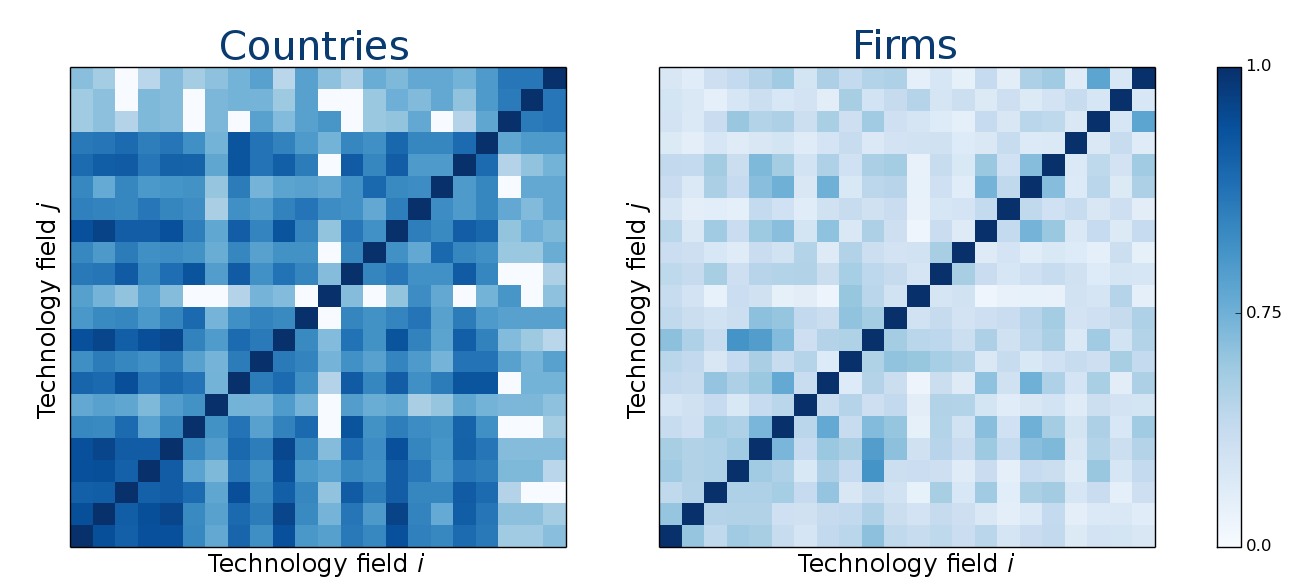}
\caption[\textbf{The structure of the relatedness matrix B}]{\textbf{The structure of the relatedness matrix B} changes if it is built from data aggregated at different geographical levels. The left panel represents $\tilde B$ computed from an aggregated version of $M$ -- say $\tilde M$ -- in which rows not longer index individual firms,  but rather the nations in which such firms reside. The right panel represents $B$ from equation \ref{eq:zac_firm_rel}, the relatedness measure used throughout the paper. The same technological codes are much more clustered in the former case than in the latter.}
\label{fig:bScale}
\end{figure}

However, there is no constraint on the geographical aggregation of the matrix from which $B$ is defined.
It is in fact possible to substitute  $M$ with another matrix in which firms are aggregated based on the country or region in which they have seat and use it to compute a new relatedness matrix $\tilde B$, in which the starting geographical aggregation can be arbitrarily coarse.\par

Figure \ref{fig:bScale} shows that indeed the geographical scale used to define the global relatedness of technological fields has a deep influence on the observed global relatedness between technologies.
In fact, there is a striking difference between the right and the left panel, which represent respectively $B$ from equation \ref{eq:zac_firm_rel} and $\tilde B$ computed from $\tilde M$. We point out that we used the IPC classification to order rows and columns. Notice that the same technological codes are much more clustered in the latter case (country-based aggregation) than in the former (in which we work at the firm level).
This finding is not surprising, given that defining technological relatedness at the national level means considering the technological portfolios of extremely differentiated entities,
which, by definition, can explore a much larger set of combinations and hence better highlight the true global relatedness structure between technological fields.
On the other hand, as large as individual companies can be, they are necessarily constrained in the breadth of their output basket and their reference market, and as a consequence they will also be limited to the development of combinations of technological capabilities needed to effectively produce that relatively narrow set of goods and services.
The question however remains as to how different definitions of $B$ affect the correlation between $\Gamma$ and specific firm characteristics and, eventually, whether an optimal geographical scale exists at which to define the global technological relatedness matrix.

\begin{table}[htb]
  \centering 
  \begin{tabular}{ r c c c c }
  VARIABLES   & (  0 )     & (  1 )     & (  2 )     & (  3 )      \\ 
  \hline			
  &&&\\
         Size & 0.082*** & 0.081*** &            & 0.082***  \\
              & \footnotesize{(0.056)}  & \footnotesize{(0.009)}  &            & \footnotesize{(0.009)}   \\
              &&&\\
  Diversification & -0.047   &            &            & 0.068***  \\
              & \footnotesize{(0.071)}  &            &            & \footnotesize{(0.010)}   \\
              &&&\\
Coherent Div. & 0.148**  & 0.089*** & 0.121*** &             \\
              & \footnotesize{(0.071)}  & \footnotesize{(0.013)}  & \footnotesize{(0.013)}  &             \\
              &&&\\
  \hline 
  &&&\\
        $R^2$ & 0.057    & 0.056    & 0.031    & 0.055     \\
  \end{tabular}
  \caption{Regressions of labor productivity against country-level coherent diversification, diversification, and size}
  \label{tab:reg_country}
\end{table}

Table \ref{tab:reg_country}, which summarizes the regression of labor productivity against coherent diversification measured based on $\tilde B$, provides evidence in line with the hypothesis that geographical aggregation actually has an influence and that measuring diversification at the company level is a more suited starting point to measure firm-level coherent diversification with respect to a more aggregated definition.
Notice that, as in table \ref{tab:reg}, coherent diversification is statistically significant in all settings and also explains productivity better than diversification.
However, the explanatory power as measured by the $R^2$ is sensibly lower in the regressions of table \ref{tab:reg_country}, suggesting that coherent diversification does not work as well if defined at a coarser geographical scale.
This finding has clear implications for those studies that apply country-based measures such as \cite{hidalgo2007product} and \cite{zaccaria2014related} directly to firm-level data.
\section{Conclusions} \label{sec:conclusioni}

In this work we have presented a quantitative assessment of the relationship between corporate technological portfolios and their performance. The idea is that successful companies shape their patenting activity on the basis of well defined production lines, and that this strategic behavior can be traced by looking at corporate technological portfolios. In particular, we introduce a methodology to reconstruct an estimate of both the size and number of the coherent blocks of knowledge a firm owns, and we show that their average size is correlated with firms performance.

From a practical point of view, we have used a database of about 70 thousand firms, including their patenting activity, in order to define a bipartite companies-technologies network. A link is present if a firm is active in a given technological field, as reported by codes in their submitted patents. Then, we have built a monopartite network of technological codes by applying a measure of relatedness initially conceived to uncover the common capabilities that countries should have to export a pair of products. This network can be used to assess the relative integration of technological activities within a firm. The result is the so called coherent diversification, a weighted average of the relatedness of a firm’s technologies, which can be seen as a proxy of the average size (in terms of technological fields) of a firms' coherent blocks of knowledge. We have found that the coherent diversification explains firms’ performance, as measured by labor productivity, in a statistically significant way, and even if 
standard diversification and size are used as controls. This finding has remarkable practical consequences: for instance, it points out that coherent diversification, and not diversification by itself, should be taken into account in merging or acquisitions among companies. Finally, we have presented a comparative analysis of the structure of the technological space when countries, and not firms, are
considered as patenting entities. We have found that a better explanatory power of firms’ performance can be obtained if the aggregation is made at firm level, which therefore represents a more representative scale for this kind of studies.

This work opens up a number of possible further studies. For instance, in our analysis the production lines represent a hidden layer that can be proxied by pinpointing coherent blocks in corporate technological portfolios. When one analyzes directly products, these blocks should clearly emerge, giving rise to well defined clusters possibly in agreement with the standard classification - while this could be not true for technological fields. The study of the different clustering behavior of product and technologies will be the subject of a forthcoming paper. 

\vfill
\pagebreak
\renewcommand\bibname{References}
\bibliography{../npz_ref}

\end{document}